\documentclass{mn2e}
\usepackage{graphicx}

\title{The unusual supernova remnant surrounding the ultraluminous
X-ray source IC 342 X-1}

\author[T.\,P. Roberts et al.]
{T.\,P. Roberts$^{1,*}$, M.\,R. Goad$^2$, M.\,J. Ward$^1$ \&
R.\,S. Warwick$^1$\\ $^1$ X-ray and Observational Astronomy Group,
Dept. of Physics \& Astronomy, University of Leicester, University
Road, Leicester, LE1 7RH\\ $^2$ Dept. of Physics \& Astronomy,
University of Southampton, Highfield, Southampton, Hants., SO17 1BJ\\
$^*$E-mail: tro@star.le.ac.uk}
\date{}

\def\ro{{\it ROSAT~\/}}
\def\asca{{\it ASCA~\/}}
\def\ein{{\it Einstein~\/}}
\def\xmm{{\it XMM-Newton~\/}}

\def\chan{{\it Chandra~\/}}

\def\ergcms{{\rm ~erg~cm^{-2}~s^{-1}}}

\def\ergsec{{\rm ~erg~s^{-1}}}

\def\H0{{\rm ~km~s^{-1}~Mpc^{-1}}}

\def\la{\mathrel{\hbox{\rlap{\hbox{\lower4pt\hbox{$\sim$}}}{\raise2pt\hbox{$<$}}}}}
\def\ga{\mathrel{\hbox{\rlap{\hbox{\lower4pt\hbox{$\sim$}}}{\raise2pt\hbox{$>$}}}}}

\def\d25{D$_{25}$}

\def\Ha{{H$\alpha$}}

\def\hii{H{\small II}$~$}

\def\.25{0.25 keV\thinspace}
\def\lx{L$_{\rm X}$}

\begin{document}

\maketitle

\begin{abstract}
We report the results of an observation of a large diameter (110 pc)
supernova remnant (SNR) found to encircle the position of the
ultraluminous X-ray source (ULX) IC 342 X-1.  The inferred initial
energy input to the SNR is at least 2 -- 3 times greater than the
canonical energy for an ``ordinary'' supernova remnant.  Two regions
on the inside of the shell are bright in [O{\small III}] $\lambda
5007$ emission, possibly as the result of X-ray photoionization by the
ULX.  If this is the case, then the morphology of this nebulosity
implies that the X-ray emission from the ULX is anisotropic.  The
presence of the ULX, most probably a black hole X-ray binary, within
an unusually energetic supernova remnant suggests that we may be
observing the aftermath of a gamma-ray burst, though other origins for
the energetic nebula are discussed.
\end{abstract}


\section{Introduction}

Ultraluminous X-ray Sources (ULX) are the most luminous point-like
extra-nuclear X-ray sources found in nearby galaxies, with observed
X-ray luminosities in excess of $10^{39} \ergsec$.  Whilst some recent
supernovae can appear as ULX (e.g. SN 1986J, Bregman \& Pildis 1992;
SN 1979C, Immler et al. 1998), the majority of ULX are believed to be
accreting systems, and indeed \asca studies have shown that many ULX
display the characteristics of accreting black holes (e.g. Makishima
et al. 2000; Mizuno, Kubota \& Makishima 2001).  Several competing
models currently provide plausible physical descriptions of these
systems, and explain how they apparently reach, and in many cases
greatly exceed, the Eddington luminosity for a 10 M$_{\odot}$ black
hole.  These include accretion onto a new class of $10^2 - 10^5$
M$_{\odot}$ intermediate-mass black holes (e.g. Colbert \& Mushotzky
1999; Miller \& Hamilton 2002), possible examples of which were
recently inferred to be present in the globular clusters G1 and M15
(Gebhardt, Rich \& Ho 2002 and references therein); truly
super-Eddington X-ray emission from ``ordinary'' X-ray binaries
(Begelman 2002); and anisotropic emission from X-ray binaries (King et
al. 2001), perhaps relativistically beamed if we are looking down the
jets of microquasars (e.g. K{\"o}rding, Falcke \& Markoff 2002).
Current observations do not rule out any of these models, but they do
suggest a heterogeneous population, with ULX associated with both the
nascent stellar populations found in star forming regions, and the old
stellar population found in elliptical galaxies (c.f. Roberts et
al. 2002; Sarazin, Irwin \& Bregman 2001 - see King 2002 for further
discussion).

\begin{figure*}
\centering
\includegraphics[width=4.2cm,angle=270]{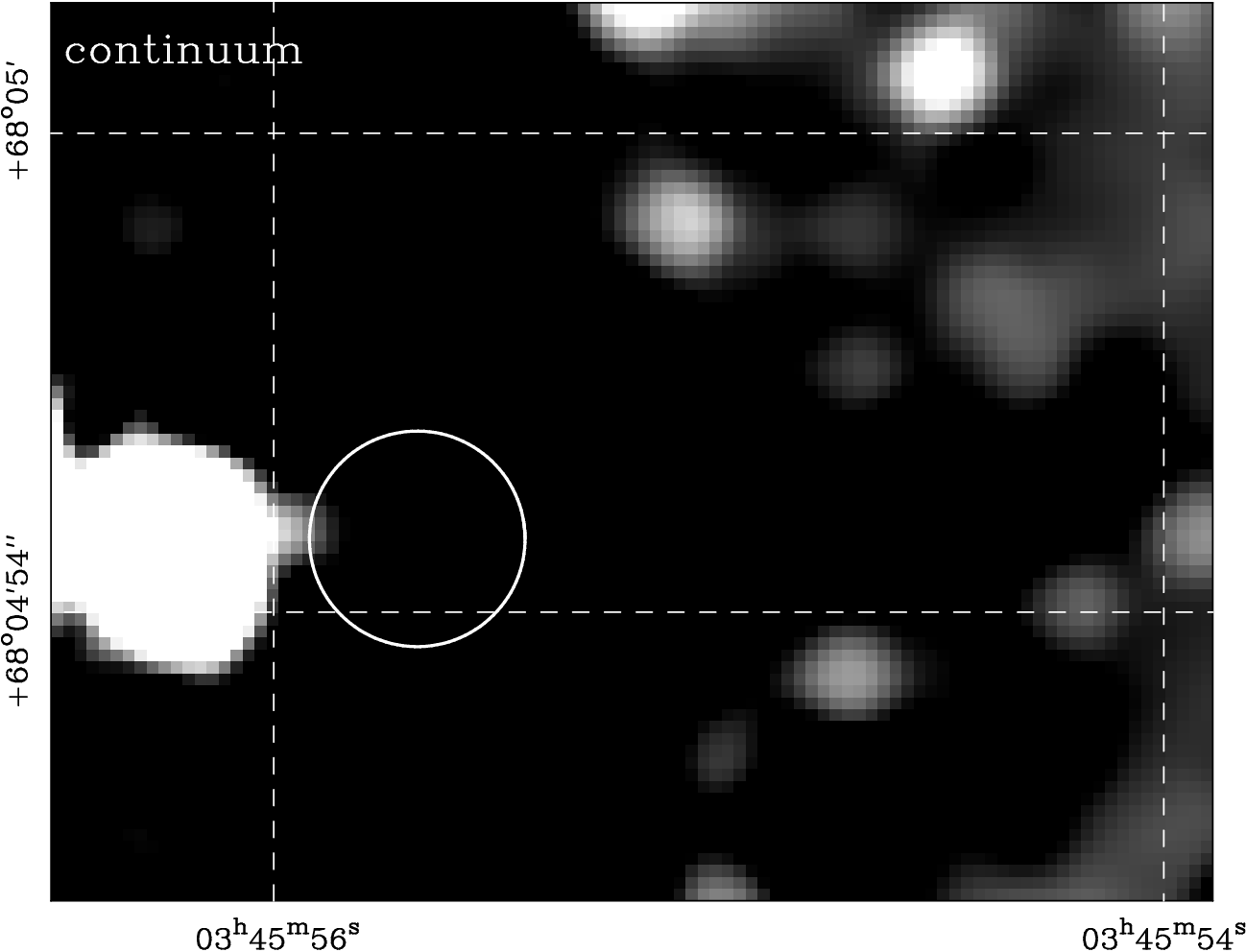}
\includegraphics[width=4.2cm,angle=270]{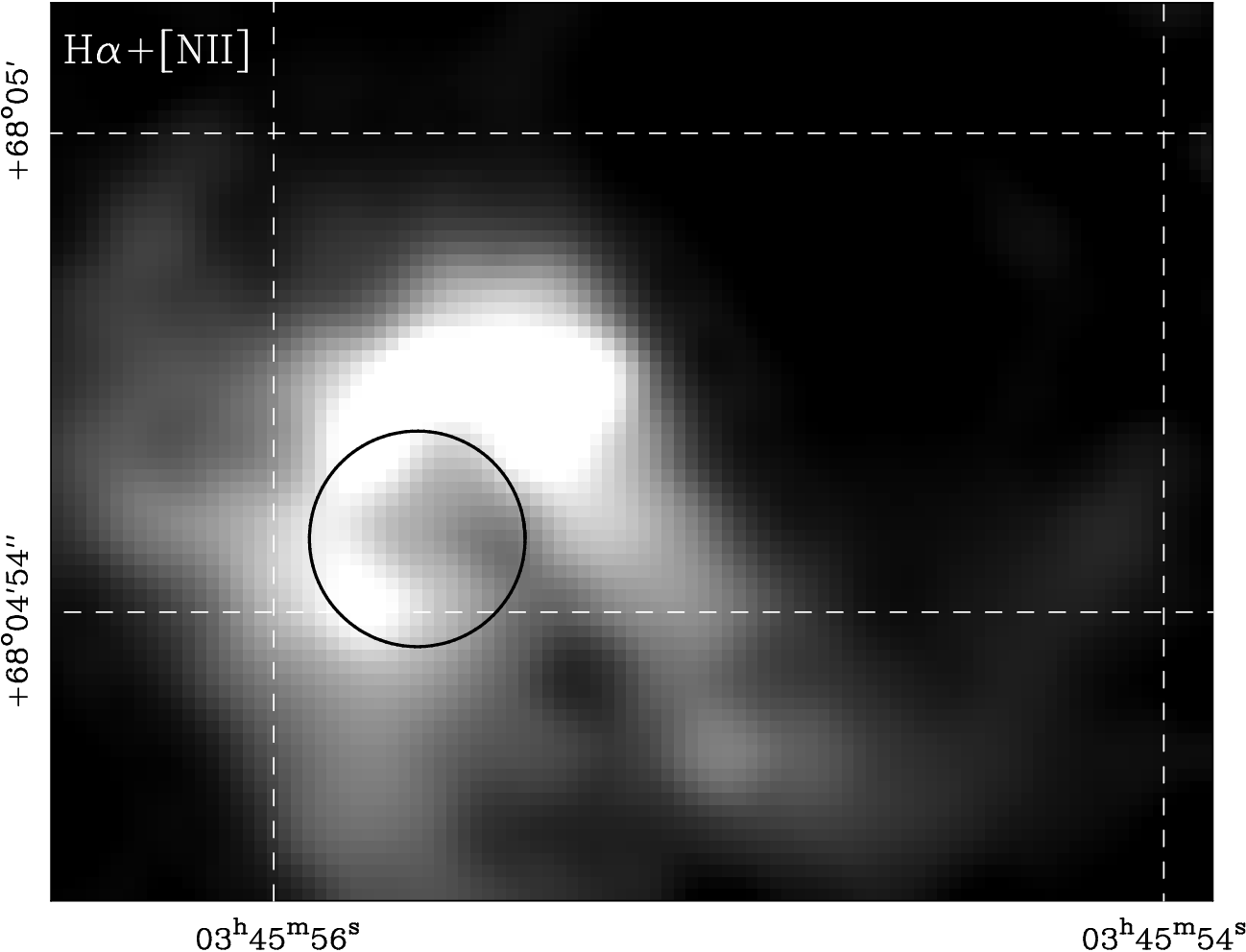}
\includegraphics[width=4.2cm,angle=270]{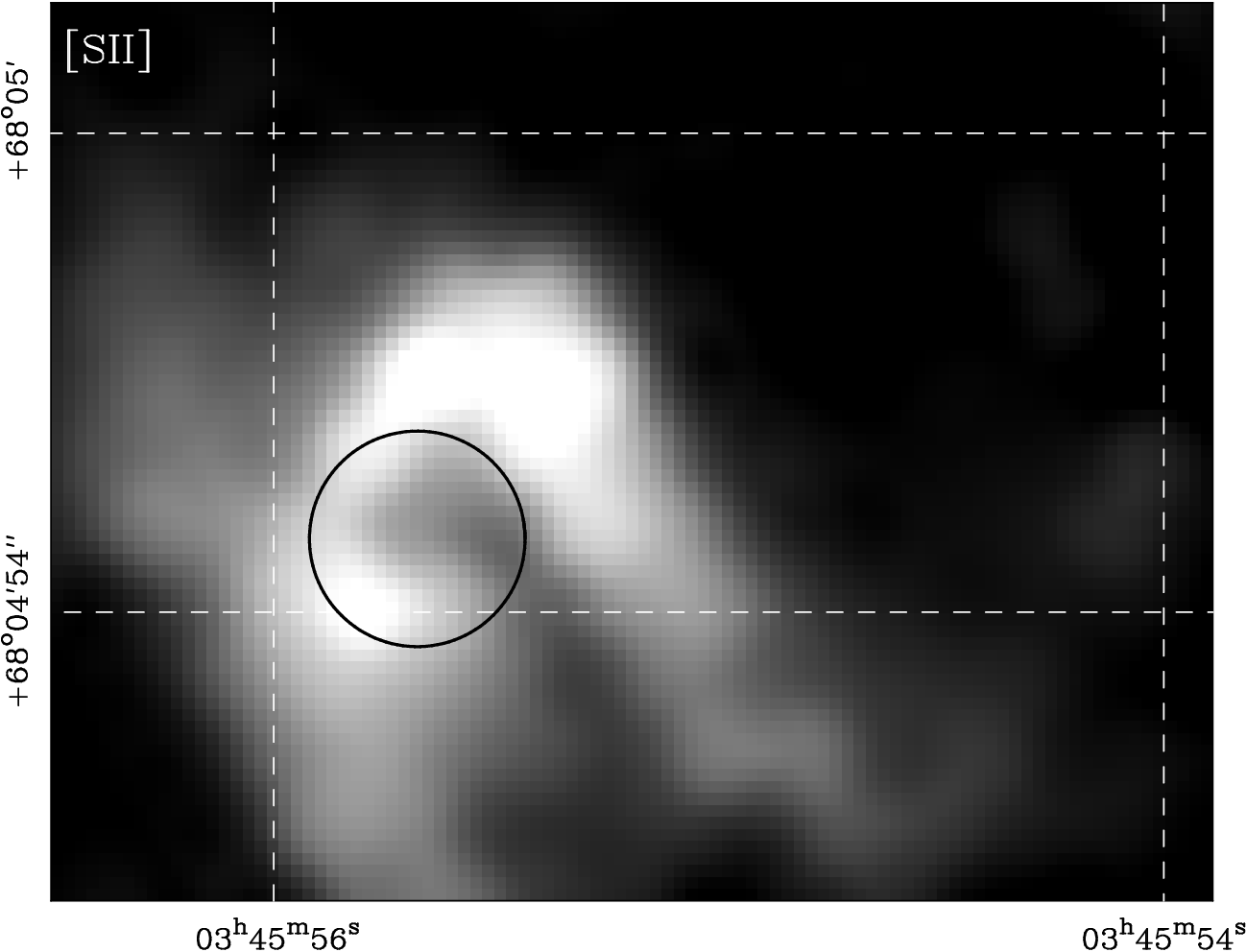}
\caption{INTEGRAL images of the IC 342 X-1 field.  The greyscale is in
units of observed surface brightness, and ranges between 0 (black) and
$2.5 \times 10^{-17} \ergcms \rm ~\AA^{-1}~arcsec^{-2}$ (white) for
the 5300 - 5500 \AA~ continuum and [S{\small II}] images, and 0 to $5
\times 10^{-17} \ergcms \rm ~\AA^{-1}~arcsec^{-2}$ for the \Ha
$+$[N{\small II}] image.  The circle represents the uncertainty in the
position of IC 342 X-1.}
\label{snr}
\end{figure*}

IC 342 X-1 is one of the nearest and most comprehensively-studied
ULX\footnote{The precise distance to IC 342 remains quite uncertain,
due to high foreground obscuration.  Previous X-ray analyses assume
distances of $\sim 4$ Mpc, consistent with the distance of 3.9 Mpc
quoted by Tully (1988), which is itself supported by a later analysis
placing the IC 342/Maffei group at a distance $3.6 \pm 0.5$ Mpc
(Krismer, Tully \& Gioia 1995).  However, other studies have placed
the galaxy much closer.  For example, photometry of supergiant stars
in IC 342 imply a distance of only 2.1 Mpc (Karachentsev \& Tikhonov
1993).  In this paper we assume a distance of 3.9 Mpc.}.  It was
originally detected in an \ein IPC observation of IC 342 with a 0.2 --
4 keV X-ray luminosity of $3.0 \times 10^{39} \ergsec$ (Fabbiano \&
Trinchieri 1987).  A subsequent \ro observation also detected the
source in an ultraluminous state (Bregman, Cox \& Tomisaka 1993;
Roberts \& Warwick 2000).  An \asca observation obtained in September
1993 showed IC 342 X-1 to be in a very luminous state (\lx $~> 1.5
\times 10^{40} \ergsec$, 0.5 - 10 keV), in which it displayed large
amplitude, short timescale ($\sim 1000$ s) variability (Okada et
al. 1997), and possessed an X-ray spectral form characteristic of
black hole accretion disc spectra in the high/soft state (Makishima et
al. 2000).  Conversely, a deep follow-up \asca observation in February
2000 showed the ULX to have dimmed to $6 \times 10^{39} \ergsec$ (0.5
- 10 keV) and undergone a spectral transition into a low/hard state,
similar behaviour to that observed in many Galactic and Magellanic
black hole binary systems (Kubota et al. 2001; Mizuno, Kubota \&
Makishima 2001).  An alternate view is that IC 342 X-1 could have been
in the very high/anomalous state during this observation, which is
also seen in many Galactic black hole binaries (Kubota, Done \&
Makishima 2002).  IC 342 X-1 is therefore one of the best candidates
for a {\it bona fide\/} black hole X-ray binary ULX.

Multi-wavelength studies are crucial to understanding the nature of
ULX, through the identification of potential counterparts and the
investigation of their immediate environment.  Several ULX have been
shown to be associated with structures apparent in other bands.  For
example, a star forming complex on the edge of NGC 4559 is host to the
luminous ULX NGC 4559 X-7 (Vogler, Pietsch \& Bertoldi 1997), and M81
X-9 is located within a shock-heated giant nebula, possibly a
supershell formed by one very energetic supernova, or multiple
supernovae (Miller 1995; Wang 2002).  However, the precise
identification of possible counterparts, or the exact location of ULX
within larger structures, has only been possible in the new era of
sub-arcsecond X-ray astrometry provided by {\it Chandra\/}.  This has
led to the first identification of possible stellar counterparts to
ULX, in particular with the identification of three young ($< 10$ Myr
old) compact stellar clusters associated with NGC 5204 X-1 (Roberts et
al. 2001; Goad et al. 2002), the coincidence of a ULX in NGC 4565 with
a faint globular cluster (Wu et al. 2002), and the identification of
an O-star counterpart to M81 X-6 ($\equiv$ NGC 3031 X-11; Liu, Bregman
\& Seitzer 2002).

The presence of optical emission-line nebulae at, or close to, the
positions of several ULX has recently been reported in a conference
proceedings paper by Pakull \& Mirioni (2003a; hereafter PM03).  This
includes the identification of a nebula coincident with the \ro HRI
position of IC 342 X-1, that they christen the ``tooth'' nebula on the
basis of its morphology.  Their optical spectroscopy shows the nebula
to display ``extreme SNR-like emission line ratios: [S{\small
II}]/\Ha~ = 1.2 and [O{\small I}]$\lambda 6300$/\Ha~ = 0.4''.  In this
letter we report new X-ray and optical observations that locate the
suspected black hole X-ray binary ULX IC 342 X-1 at the heart of this
unusual nebula.

\section{Observations \& results}

The optical data were obtained on the night of 2001 February 1 using
the INTEGRAL field spectrograph on the William Herschel Telescope
(Arribas et al. 1998).  A summary of the instrumental set-up is given
in Roberts et al. (2001) and the data analysis will be detailed in
Roberts et al. (in preparation).  In short, the INTEGRAL data provided
us with spectroscopy in 189 fibres and, through the relative fibre
positions, simultaneous imaging within a $16.5'' \times 12.3''$
field-of-view.  The IC 342 X-1 follow-up observation was targetted on
$03^h45^m55.2^s, +68^{\circ}04'56''$, which is the \ro HRI position
after applying an astrometric correction, calculated from the field
X-ray source positions and their possible optical counterparts in the
USNO catalogue\footnote{All co-ordinates in this letter are quoted in
epoch J2000.}.

A new 9.9 ks \chan X-ray observation of IC 342 X-1 was obtained on
2002 April 29.  The ULX was positioned on the ACIS-S3 chip, which was
operated in the 1/8 sub-array mode to mitigate the anticipated effects
of pile-up.  Initial analysis of the cleaned data (reduced using
{\small CIAO} v2.2) detected IC 342 X-1 at a position
$03^h45^m55.68^s, +68^{\circ}04'54.9''$.  An investigation of the
radial profile of the source indicates that it is point-like at the
$0.5''$ resolution of {\it Chandra\/}.  Unfortunately, only one other
source was detected on the S3 chip (at $03^h46^m32.90^s,
+68^{\circ}03'56.0''$), and this faint source does not have an optical
counterpart, hence a direct check of the astrometry was not possible.
We therefore adopt the \chan position for IC 342 X-1, and assume a
conservative accuracy of $\pm 1''$ for the X-ray position (c.f. Goad
et al. 2002).  We defer further analysis of the X-ray emission of IC
342 X-1 to a future paper; for the remainder of this letter we
concentrate upon the interpretation of the INTEGRAL data in the
context of the precise determination of the X-ray position.

Several optical continuum sources and an emission-line nebula are
evident in the INTEGRAL field-of-view.  We show a continuum image,
plus continuum-subtracted emission-line images in \Ha $+$[N{\small
II}] and [S{\small II}] in Figure~\ref{snr}.  The nebula is
characterised by a shell-like morphology, and has a high [S{\small
II}]/\Ha~ emission-line flux ratio of $\sim 1.1$ throughout its
extent, hence we interpret it as a supernova remnant (SNR), consistent
with the comments of PM03 (c.f. Matonick \& Fesen 1997 and references
therein).  This remnant was not reported in a previous search for such
objects in IC 342 by D'Odorico, Dopita \& Benvenuti (1980).  In
Figure~\ref{snr} we illustrate the uncertainty in the position of IC
342 X-1 using a circle of radius $1.5''$; this is a conservative
estimate based on the intrinsic \chan astrometry error discussed
above, with an additional contribution of $\pm 1''$ corresponding to
the uncertainty in the INTEGRAL astrometry.  The ULX is clearly not
associated with the bright continuum source at the eastern edge of the
image, but is in fact positionally coincident with the central regions
of the supernova remnant.  This raises the intriguing possibility that
the ULX may be physically related to the supernova remnant.

\begin{figure}
\centering
\includegraphics[width=4.5cm,angle=270]{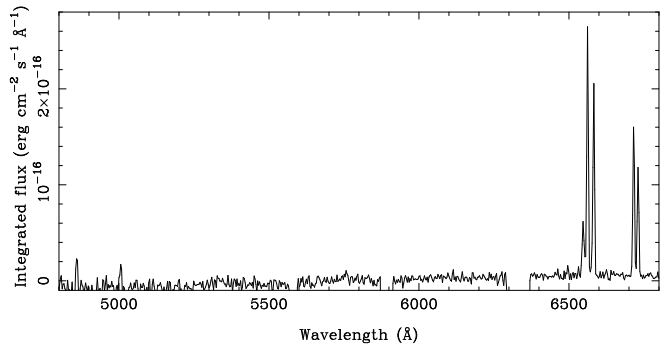}
\caption{The total flux-calibrated INTEGRAL spectrum of the SNR.  This 
spectrum is {\bf not} corrected for Galactic extinction.  Small
portions of the spectrum are excluded around the Na-D, He{\small I}
and [O{\small I}] sky lines to remove residual contamination.}
\label{snrspec}
\end{figure}

In Figure~\ref{snrspec} we show the INTEGRAL spectrum of the supernova
remnant, integrated over the whole spatial extent of the nebulosity
(58 out of 189 fibres) within the field-of-view.  It is immediately
obvious that the spectrum suffers from a high degree of reddening; in
fact the line-of-sight to IC 342 has an extinction of $E(B-V) = 0.6$,
a consequence of its low Galactic latitude ($b \sim 10^{\circ}$).
This severely limits the sensitivity at the blue end of our spectrum.
The residual contamination from sky lines is removed from
Figure~\ref{snrspec}.  Large positive residuals were found in the data
at the positions of the [O{\small I}] $\lambda 6300$ and $\lambda
6363$ lines, but the relatively low spectral resolution ($\sim 6$ \AA)
of the data and the added complication of excluding spatially-variable
sky lines from fibre data implies that any physical measurements from
these residuals would be very uncertain.  We therefore remove them
from our analysis.  The reddening-corrected integrated fluxes of the
diagnostically important emission lines are given in
Table~\ref{lines}.

\begin{table}
\centering
\caption{Integrated emission-line fluxes, corrected for
reddening.}
\begin{tabular}{lcc}\hline
Species	& $\lambda$ (\AA)	& Flux ($\times 10^{-15} \ergcms$) \\\hline
H$\beta$	& 4861	& $1.7 \pm 0.5$ \\
$[$O{\small III}]	& 4959	& $0.4 \pm 0.2$ \\
		& 5007	& $1.2 \pm 0.2$ \\
$[$N{\small II}]	& 5755	& $0.4 \pm 0.2$ \\
		& 6548	& $1.6 \pm 0.2$ \\
		& 6583	& $4.6 \pm 0.2$ \\
\Ha		& 6563	& $5.8 \pm 0.1$ \\
$[$S{\small II}]	& 6717	& $3.7 \pm 0.1$ \\
		& 6731	& $2.6 \pm 0.1$ \\\hline
\end{tabular}
\label{lines}
\end{table}

The inferred characteristics of the supernova remnant are listed in
Table~\ref{plasma}.  Both the radius of the bright northern part of
the supernova shell\footnote{We conservatively use the brightest (and
hence the densest) part of the shell, since it provides the best
emission line constraints.  We note that the lower surface brightness
extension of the nebula, particularly to the south of the shell
structure, may be due to density gradients in the ambient ISM allowing
faster expansion in this direction.}, $R_{neb}$, and the velocity of
the material in the shell, $V_s$, are directly derived from our
INTEGRAL observations ($V_s$ is a 3$\sigma$ upper limit on the
velocity of the \Ha~ emitting region, determined from a comparison of
the \Ha~ line FWHM with that of the sky lines).  However, the low
[O{\small III}]/H$\beta$ ratio of $\sim 1$ suggests that the actual
velocity may indeed be lower, probably $< 100$ km s$^{-1}$ (Dopita et
al. 1984).  To derive the age of the SNR and its initial explosion
energy we follow the method shown in PM03, who use the pressure-driven
snowplough phase equations of Cioffi, McKee \& Bertschinger (1988) to
describe the current expansion of an old, very large SNR.  As with
PM03, this relies on an estimate of the ambient ISM density around the
supernova progenitor as determined from the total radiative H$\beta$
flux of the nebula (in our case a luminosity of $3 \times 10^{36}
\ergsec$), using the equations of Dopita \& Sutherland (1995).  The
limits derived from these equations are shown in Table~\ref{plasma}.
We also show limits on the electron density and temperature in the
shocked region calculated from the [S{\small II}] $\lambda
6731/\lambda 6717$ (density) and [N{\small II}] ($\lambda 6548+\lambda
6583)/\lambda 5755$ (temperature) ratios, as per Osterbrock (1989).
Table~\ref{plasma} indicates that the supernova remnant is unusually
large, having a projected diameter of at least 110 pc.  For
comparison, Matonick \& Fesen (1997) argue that a typical single
supernova remnant will not remain visible once it has expanded beyond
a diameter of 100 pc, for a canonical energy input of 10$^{51}$ erg.
One way in which a single supernova remnant could reach the observed
size is if the initial explosion energy exceeds 10$^{51}$ erg, which
appears to be the case for this nebula since we calculate a lower
limit on the initial energy input of $\sim 2 \times 10^{51}$ erg.  We
discuss this possibility further in Section 4.

\begin{table}
\centering
\caption{The properties of the supernova remnant.}
\begin{tabular}{@{}llc@{}} \hline 
Radius & $R_{neb}$ & 55~pc \\
Shell velocity & $V_s$ & $<$~180~km~s$^{-1}$ \\
Age & $\tau_{neb}$ & $>$~92000~yr \\ 
Initial energy ($\times 10^{51}$ erg) & $E_{51}$ & $> 2$ \\
Ambient ISM density	& $n_0$	& $> 0.12$ cm$^{-3}$\\
Electron density $^a$ & $N_{e}$ & $<$~40~cm$^{-3}$\\ 
Electron temperature $^b$ & $T_{e}$  & $< 3\times 10^{4}$~K\\ \hline
\end{tabular}
\begin{tabular}{l}
Notes: $^a$ from [S{\small II}] $\lambda 6731/\lambda 6717 = 1.4 \pm
0.1$.\\ $^b$ From [N{\small II}] ($\lambda 6548+\lambda 6583)/\lambda
5755 < 15.5$.
\end{tabular}
\label{plasma}
\end{table}

\section{Evidence for an X-ray ionized nebula inside the supernova
remnant?}

\begin{figure}
\centering
\includegraphics[width=6cm,angle=270]{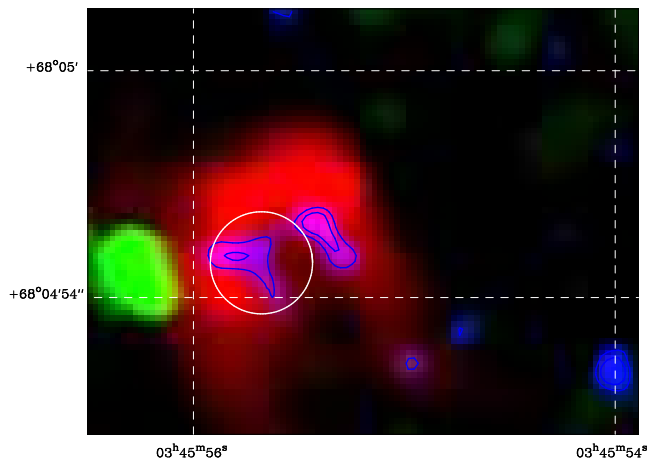}
\caption{The unusual ionization structure of the IC 342 X-1 SNR.  The
three colours represent 5300 -- 5500 \AA~ continuum emission (green),
continuum-subtracted H$\alpha$+[N{\small II}] emission (red), and
continuum-subtracted [O III] emission (blue).  The [O III] emission is
highlighted by blue contours showing the 3 and 3.5$\sigma$ confidence
levels above the background noise.  The uncertainty in the position of
IC 342 X-1 is shown by the circle.}
\label{3col}
\end{figure}

A remarkable spatial feature of this supernova remnant is revealed
upon closer inspection of the continuum-subtracted emission-line
images.  Whilst the \Ha, [N{\small II}] and [S{\small II}] images all
trace the same shell structure, an image created from the faint
[O{\small III}] line shows a distinctly different morphology,
illustrated in Figure~\ref{3col}.  The faint [O{\small III}] emission
appears concentrated on the inside of the supernova remnant, and
traces the inside of the shell, both within the positional error
circle of the ULX and to the north-west of this position.  However,
this [O{\small III}] nebula does not appear to follow the entire
structure of the inside of the shell, with no emission detected along
the inside of the northern part of the shell.  The presence of
enhanced [O{\small III}] emission on the inner-edge of the shell is
confirmed by extracting spectra from individual fibres in both the
[O{\small III}] region and to the north of this position, which
confirm a higher average [O{\small III}]/H$\beta$ ratio in the former
(1.0 vs 0.5 in the respective reddening-corrected
ratios)\footnote{These [O{\small III}]/H$\beta$ ratios are not
indicative of high excitation regions.  However, given the spatial
resolution of $\sim 20$ pc per fibre, this of course does not preclude
smaller parsec-scale high-excitation regions within the nebula.}.  It
therefore appears that there is a concentration of highly-excited
oxygen located on the inside of the much larger supernova remnant.
Furthermore, given that the shock front passed through this region of
the nebula at least half the SNR lifetime ago (assuming a physically
unrealistic constant expansion velocity), and that the [O{\small III}]
recombination time is much shorter than the age of the nebula
($\tau_{rec} > 150$ years for $N_e < 40$ cm$^{-3})$\footnote{We note
that $N_e$ may be a factor 100 smaller, and thus $\tau_{rec}$ 100
times larger, if the temperature exceeds 15000 K.  This is still much
less than the age of the remnant.}, it is possible that a process
other than the supernova blast wave may be energising the [O{\small
III}] emission.  If this is so, what could be producing it?

One possibility is that this high-excitation emission is produced by
the ULX.  In particular, it could potentially be photoionized by the
$> 10^{39} \ergsec$ X-ray flux to produce an ``X-ray ionized nebula''
(XIN).  Relatively few examples of this phenomenon are currently
known.  Two examples are the XIN associated with the high-mass X-ray
binary LMC X-1 (Pakull \& Angebault 1986) and the supersoft X-ray
source CAL 83 (Remillard, Rappaport \& Macri 1995), both in the Large
Magellanic Cloud.  Also, an XIN was recently discovered associated
with the ULX in Holmberg II (PM03).  On first appearances, our
[O{\small III}] nebula bears remarkable similarities to the [O{\small
III}] nebula surrounding CAL 83, which is also located on the inside
of an incomplete \Ha~ shell.  However, our observed [O{\small
III}]/H$\beta$ flux ratio is much smaller than that observed around
CAL 83, which has a ratio [O{\small III}]/H$\beta = 15 \pm 2$,
indicative of a very high excitation state (though see footnote $^4$).
This difference may perhaps be a result of the very different X-ray
spectra of CAL 83 and IC 342 X-1.  CAL 83 is a classic supersoft X-ray
source, with a very soft X-ray spectrum that produces a large flux of
UV and very soft X-ray photons, that are ideally suited to
photoionizing its immediate environment (Remillard et al. 1995),
whereas IC 342 X-1 displays the much harder X-ray spectrum
characteristic of a black hole X-ray binary.  Also, in the case of CAL
83 the surrounding medium is very different, with the XIN being
produced in the ISM in the vicinity of the X-ray source, as opposed to
in the inner parts of a supernova shell.

It is therefore informative to consider other possible signatures of
X-ray ionization.  The XIN around LMC X-1 and the Holmberg II ULX were
primarily identified by the presence of relatively faint He{\small II}
$\lambda 4686$ lines in their optical spectra.  Unfortunately, the
high extinction towards IC 342 means that our INTEGRAL spectrum is
dominated by noise at short wavelengths, so we are not sensitive to
faint line emission below H$\beta$.  The observational evidence for
the [O{\small III}] nebula being X-ray ionized is therefore
inconclusive, at least on the basis of the INTEGRAL data alone.

Another method of addressing whether the [O{\small III}] nebulosity is
an XIN is simply to consider whether we would expect the ULX to create
a photoionized zone consistent with the size of the nebula.  We
estimate the number of available photoionizing X-ray photons using the
best-fit to the \asca data for IC 342 X-1 in its high state, namely a
multi-colour disc blackbody model (Kubota et al. 2001).  By
extrapolation this gives a hydrogen ionizing flux of $Q = 2.3 \times
10^{48}$ photon s$^{-1}$ in the 13.6 -- 300 eV range, originating in
the ULX and not escaping the nebula (calculated in {\small XSPEC},
assuming a hydrogen density of of 0.12 atom cm$^{-3}$ in the nebula,
and that the nebula becomes essentially transparent to X-ray photons
at 300 eV).  This ionizing flux will create a Stromgren sphere in \Ha~
up to 170 pc in radius for a low density (0.12 atom cm$^{-3}, \equiv
n_0$), homogeneous medium.  However, for IC 342 X-1 we should consider
the case in which a higher density medium (the SNR shell) surrounds a
low density cavity.  Assuming that IC 342 X-1 is located in the centre
of the cavity, placing it $\sim 20$ pc from the nebula, and that the
cavity is essentially empty whilst the edge of the nebula has a
density of e.g. 3 atom cm$^{-3}$, then a similar calculation shows
that the ULX will photoionize the surface of the nebula, penetrating
up to 7 pc into the denser gas.  Of course, in reality the density of
3 atom cm$^{-3}$ is higher than the predicted post-shock
density\footnote{The density in the post-shock regions may be
estimated as $N_S = 4n_0$ (Osterbrock 1989), implying $N_S \sim 0.5$
cm$^{-3}$ in this case.}, and the edge of the nebula may not be
tightly defined, which should act to increase the physical size of the
ionized zone.  Finally, we note that following the well-known relation
L(\Ha ) $= 1.3 \times 10^{-12} Q \ergsec$ implies that a maximum \Ha~
luminosity of $\sim 3 \times 10^{36} \ergsec$ originates in the XIN,
though this probably has a large error margin due to the uncertainty
inherent in extrapolating the ULX spectrum below 300 eV, and in using
a relation that is strictly valid for stellar EUV sources.  This is
below the observed \Ha~ luminosity of $1 \times 10^{37} \ergsec$,
supporting the scenario in which an XIN sits on the inner edge of a
larger SNR which has produced the bulk of the \Ha~ excitation.  

An alternative interpretation is that the large uncertainty in the
errors could be consistent with the entire \Ha~ luminosity originating
in X-ray ionization.  It would imply the whole nebula is X-ray
ionized, and not a supernova remnant.  In this paper we interpret the
object as a supernova remnant on the basis of a high [S{\small
II}]/\Ha~ ratio implying a shock excited nebula.  We note that Pakull
\& Mirioni (2003b) suggest that X-ray ionized nebulae may contain an
extended warm ($T_e = 10^4$K) low-ionization region which would excite
strong lines of near-neutral species such as [O{\small I}] and
[S{\small II}], thereby mimicking the emission-line ratios of a
shock-excited nebula.  Further work to confirm this model would be of
great importance.


The above discussion assumes that the X-ray emission of the ULX is
isotropic.  In this case the morphology of the photoionized nebula
should reflect the distribution of the surrounding material.  However,
the discrepancy between the bi-modal [O{\small III}] distribution and
the half-shell distribution in the supernova remnant suggests that the
nebula is not seeing an isotropic photoionizing source.  This implies
that the X-ray emission of the ULX is anisotropic, as has been
suggested to explain the apparent super-Eddington luminosities of ULX
(King et al. 2001).

A potential alternative source of ionization for the [O{\small III}]
nebulosity is one or more young stars.  The presence of the supernova
remnant and the ULX suggest that young stars exist in this region of
IC 342 (a link between many ULX and young stellar populations appears
increasingly likely; see e.g. Roberts et al. 2002).  The total
[O{\small III}] luminosity in the central regions of the supernova
remnant (L$_{\rm [OIII], \lambda 5007} \sim 9 \times 10^{35} \ergsec$)
is consistent with the observed luminosities of nearby \hii regions in
our Galaxy.  Indeed, the Orion nebula has a luminosity of L$_{\rm
[OIII], \lambda 5007} \sim 2 \times 10^{37} \ergsec$ (Pogge, Owen \&
Atwood 1992), with the ionisation originating in a single O7V star,
$\theta^1$ Ori C, possessing an ionising flux of $Q = 10^{49 - 50}$
photon s$^{-1}$ (Ferland 2001).  Curiously, if ULX are black-hole
X-ray binaries containing a young secondary star (c.f. King et
al. 2001), then it is plausible that the [O{\small III}] excitation
might originate in the secondary star in the ULX system, though we
note that the morphology of the [O{\small III}] nebula might argue
against a single stellar excitation source.  The presence of one or
more O-stars is not ruled out by our continuum images, which only
probe down to M$_V \sim -9$ in IC 342 due, in part, to the heavy
line-of-sight reddening.  Deeper optical continuum observations of the
environment of IC 342 X-1 will provide the crucial constraints on
the presence of young stars.

\section{A black hole formed in a hypernova explosion?}

As alluded to in Section 2, the idea that the ULX may be causally
related to an unusually energetic supernova remnant is an intriguing
one.  The case for a physical link (as opposed to a line-of-sight
coincidence) between the two is made stronger if we can attribute the
[O{\small III}] nebular emission within the SNR to the ULX at its
centre.  As discussed in the introduction, this particular ULX is one
of the best candidates for a {\it bona fide\/} black hole X-ray
binary, hence we have a scenario in which a black hole system sits
inside an energetic supernova remnant.  This appears to satisfy the
conditions for the aftermath of a gamma-ray burst, in which the
collapse of a massive star to a black hole triggers a supernova
explosion with an unusually high input energy, i.e. a ``hypernova''
(e.g. Paczy{\'n}ski 1998; Fryer \& Woosley 1998).  If so, this system
may be direct evidence that gamma-ray bursts do form black holes.

This hypothesis requires further examination.  In particular, it is
important to establish that this supernova remnant really is a
hypernova remnant (assuming even that it is a supernova remnant - see
previous section).  Following the methods adopted by PM03 we derived
an initial input energy of $E_{51} > 2$ (see Section 2), assuming a
velocity of 180 km s$^{-1}$.  If the actual velocity is $< 100$ km
s$^{-1}$, as suggested by the [O{\small III}]/H$\beta$ ratio, then
$E_{51}$ increases, though only to $> 3$.  If we accept the same
convention as Chen et al. (2002) and define a hypernova remnant as a
supernova remnant possessing an input energy in excess of $10^{52}$
erg, the supernova remnant clearly does not meet this strict criterion
(though it could be met if the actual nebular density is closer to the
value of $< 40$ cm$^{-3}$ inferred from the [S{\small II}] ratio than
the assumed density in Table~\ref{plasma}).  However this remnant, if
it originates in a single supernova explosion, still appears to be
more energetic than an ``ordinary'' supernova by a factor at least 2
-- 3.

Other scenarios for the creation of this energetic supernova remnant
should also be considered.  In particular, this remnant is
morphologically very similar to the shell surrounding the ULX M81 X-9,
albeit smaller in size (c.f. Miller 1995), which is suggested to be
the result of multiple supernovae.  A ``superbubble'' of this size
might be distinguishable from a single energetic supernova remnant by
their contrasting velocities ($\ll 100$ km s$^{-1}$ for a superbubble;
c.f. Chen et al. 2002).  In this case a minimum of only 2 -- 3
``ordinary'' supernova events occuring inside the bubble within a
relatively short time ($\sim 10^5$ years) could inject sufficient
energy to expand the bubble to its current size.  This relatively high
incidence of supernovae would require a population of young stars to
reside inside the bubble, which again should be unveiled by deeper
optical observations.

A further source of power that could expand the nebula is a jet and/or
stellar wind originating from the ULX system.  If we follow PM03, who
note that the total energy requirements and lifetime of a system
inflated by such a process would be quantitatively similar to that
from a SNR, then the energetics of the nebula imply that an average of
$\sim 10^{39} \ergsec$ would need to be injected over its lifetime.
This would require a mechanical energy of the wind similar to the
radiation energy of the ULX.  This is consistent with the average
energy input of $3 \times 10^{39} \ergsec$ estimated to be required to
inflate the W50 system in our own galaxy, probably originating in the
relativistic jets of the microquasar SS 433 (Dubner et al. 1998).
PM03 also note that mildly relativistic jets, similar to those of SS
433, could be responsible for inflating many of the nebulae in their
sample.  This scenario could therefore constitute a very credible
alternative to an energetic supernova event.

Finally, is the presence of the ULX consistent with a black hole
possibly formed only $\sim 10^5$ years ago?  If the black hole has
formed in isolation it could be accreting directly from the ISM in its
vicinity via Bondi-Hoyle accretion.  However, this accretion mode is
unlikely to produce the required X-ray luminosity, even if any
``kick'' it received on formation was sufficient to move the black
hole into the denser nebular regions (c.f. Miller \& Hamilton 2002).
It is therefore much more likely that the black hole is in an
accreting binary system.  Such a system might expect to receive a kick
on its formation in a supernova explosion.  If we assume a similar
kick to that experienced by the Galactic black hole system GRO 1655-40
of $\sim 100$ km s$^{-1}$ (Mirabel et al. 2002), then it will have
travelled no more than $\sim 10$ pc.  It should clearly still be
located close to the centre of the supernova remnant, which is
consistent with Figures~\ref{snr} \& \ref{3col}.  However, to produce
a luminous accreting binary system $\sim 10^5$ years after the black
hole was formed requires that both the primary (black hole progenitor)
and the secondary (donor) stars have very similar, large initial
masses (c.f. Verbunt \& Van den Heuvel 1995).  The formation of such a
binary system appears unlikely to occur in isolation, and may imply
the presence of a young, dense cluster similar to those seen close to
ULX in NGC 5204 (Goad et al. 2002) and the Antennae (Zezas et
al. 2002), albeit one not bright enough to be detected in our
observation of this field.

\vspace{0.2cm}

{\noindent \bf ACKNOWLEDGMENTS}

We thank the referee, Manfred Pakull, for his many useful comments
that have greatly improved this paper.  TPR gratefully acknowledges
financial support from PPARC.  TPR and MJW thank the Aspen Center for
Physics for their hospitality during the drafting of this paper.

\end{document}